\documentclass[twocolumn,showpacs,preprintnumbers,amsmath,amssymb,superscriptaddress]{revtex4}

\usepackage{graphicx}
\usepackage{dcolumn}
\usepackage{bm}

\begin{document}


\title{Ultimate on-chip quantum amplifier}

\author{O. Astafiev}
\affiliation{NEC Nano
Electronics Research Laboratories, Tsukuba, Ibaraki 305-8501, Japan}
\affiliation{RIKEN Advanced Science Institute, Tsukuba, Ibaraki 305-8501, Japan}

\author{A. A. Abdumalikov, Jr.}
\altaffiliation[On leave from ]{Physical-Technical Institute, Tashkent 100012, Uzbekistan}
\affiliation{RIKEN Advanced Science Institute, Tsukuba, Ibaraki 305-8501, Japan}

\author{A. M. Zagoskin}
\affiliation{Department of Physics, Loughborough University, Loughborough, LE11 3TU Leicestershire, UK}

\author{Yu.\ A. Pashkin}
\altaffiliation[On leave from ]{Lebedev Physical Institute, Moscow 119991, Russia} \affiliation{NEC Nano
Electronics Research Laboratories, Tsukuba, Ibaraki 305-8501, Japan}
\affiliation{RIKEN Advanced Science Institute, Tsukuba, Ibaraki 305-8501, Japan}

\author{Y. Nakamura}
\affiliation{NEC Nano Electronics Research Laboratories, Tsukuba,
Ibaraki 305-8501, Japan}
\affiliation{RIKEN Advanced Science Institute, Tsukuba, Ibaraki 305-8501, Japan}

\author{J. S. Tsai}
\affiliation{NEC Nano Electronics Research Laboratories, Tsukuba,
Ibaraki 305-8501, Japan}
\affiliation{RIKEN Advanced Science Institute, Tsukuba, Ibaraki 305-8501, Japan}

\date{\today}

\begin{abstract}
We report amplification of electromagnetic waves by a single artificial atom in open 1D space. Our three-level artificial atom -- a superconducting quantum circuit -- coupled to a transmission line presents an analog of a natural atom in open space. The system is the most fundamental quantum amplifier whose gain is limited by a spontaneous emission mechanism. The noise performance is determined by the quantum noise revealed in the spectrum of spontaneous emission, also characterized in our experiments.
\end{abstract}

\maketitle

The quantum amplifiers are actively used devices and most of them rely on natural intra-atomic or molecular transitions with almost untunable transition frequencies \cite{Scully,Silfvast}. Demonstration of amplification on a single atom or molecule in open space is possible \cite{MolAmp}, however, extremely difficult due to another common characteristic of natural atoms (molecules, quantum dots): They are relatively weakly coupled to the spatial electromagnetic waves in real experiments \cite{Gerhardt,Wrigge,Tey,Vamivakas,Muller,MolAmp}, in spite of theoretical feasibility of perfect coupling by careful matching of the spacial modes to the atom \cite{Zumofen}. An alternative approach is coupling of the atoms to a field of a high quality resonator \cite{Raimond,HoodKimble,Englund,Wallraff,Schuster,Hofheinz}, which has been successfully used to demonstrate lasing action on single natural \cite{maser,Kimble} and artificial \cite{Astafiev,Nomura} atoms. In the resonators, the atom is coupled to a single mode. On the other hand, an elementary (ultimate) quantum amplifier avoids this limitation and can be realized on a single atom strongly coupled to a continuum of electromagnetic modes of open space. The matching problem of spacial modes of electromagnetic waves can be solved by reducing space dimensionality to a 1D \cite{Shen,Chang}. Recently, the highly efficient coupling of an artificial atom to an open 1D transmission line has been achieved experimentally \cite{RF}.

We demonstrate amplification on a single three-level artificial atom coupled to a 1D transmission line. The atom is a fully controllable and tunable quantum system, with all its basic characteristics, such as energy splitting and coupling to the line, designed in accordance with our requirements. Our demonstration opens the perspective of developing a new type of on-chip quantum amplifiers and other quantum devices, capable of both reproducing the known quantum-optical phenomena and realizing the novel ones, which will use the tunability, controllability and strong coupling.

\begin{figure}[tbp]
\includegraphics[width=8.6cm]{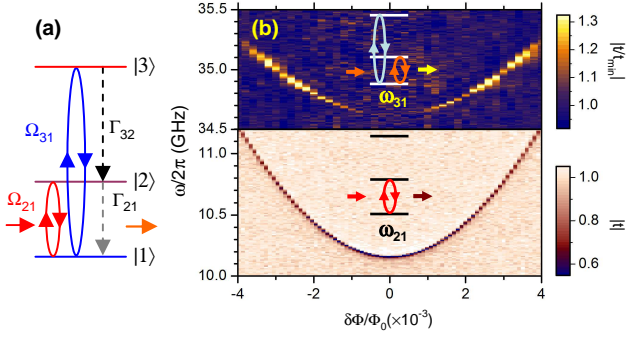}
\caption{Spectroscopy of the single artificial three-level atom.
(a) Sketch of a three-level artificial atom. Population inversion is created by pumping the atom from the ground state $|1\rangle$ to the second excited state $|3\rangle$ in the relaxation process $|3\rangle \rightarrow |2\rangle$. $\Omega_{31}$ and $\Omega_{21}$ are the pumping and probing Rabi frequencies, which are used to express the field amplitudes. (b) Spectroscopy of the single three-level atom. The lowest transition frequency $\omega_{21}$ is detected by measuring direct transmission. The higher transition frequency $\omega_{31}$ is found in the two-wave spectroscopy, and is seen as a bright line.}\label{fig:Spectroscopy}
\end{figure}

Our device is a multi-level quantum system based on the ``flux qubit" geometry \cite{Mooij} (a superconducting loop with four tunnel junctions), coupled to a 1D transmission line through the loop-line mutual inductance $M$ \cite{Abdumalikov}. We limit our consideration to the three lowest energy states of the system $|i\rangle$ ($i\: = \:1, 2, 3$) with energies $\hbar\omega_{i}$ schematically shown in Fig.~\ref{fig:Spectroscopy}(a). The device is designed in such a way that all relevant transition frequencies of the three-level system $\omega_{ij} = \omega_{i}-\omega_{j}$ ($i > j$) fall within the frequency band of our transmission line and within the  working range of our microwave sources, limited by 40~GHz. The system is strongly anharmonic, which prevents non-resonant transitions. The energies of the atomic levels are tuned by the external magnetic field. The transition frequencies $\omega_{21}$, $\omega_{32}$ and $\omega_{31}$ reach their extremal values, when the induced magnetic flux in the loop $\Phi$ equals to half a flux quantum ($\Phi_0/2$), that is $\delta\Phi = \Phi - \Phi_0/2 = 0$. Our experiment is performed at the temperature $T$ = 40 mK, low compared to the atomic energy splitting ($\hbar\omega_{ij}\gg k_BT$), which guarantees the absence of thermal excitations.

In the rotating wave approximation, the three-level system is described by the Hamiltonian
\begin{equation}\label{Eq1}
    H_a = -\hbar(\delta\omega_{21} \sigma_{22} + \delta\omega_{31} \sigma_{33}),
\end{equation}
where $\sigma_{ij} = |i\rangle\langle j|$ is the atomic projection/transition operator, $\delta\omega_{ij}=\omega_{ij}^d-\omega_{ij}$ and $|\delta\omega_{ij}| \ll \omega_{ij}$.
 The external pumping microwave fields at frequencies $\omega_{31}^d$ and $\omega_{21}^d$ couple atomic states according to the interaction Hamiltonian
\begin{equation}\label{Eq2}
    H_{\rm int} = -\hbar\Bigg[\frac{\Omega_{31}}{2}(\sigma_{31}+\sigma_{13}) + \frac{\Omega_{21}}{2}(\sigma_{21}+\sigma_{12})\Bigg],
\end{equation}
where $\hbar\Omega_{ij} = \phi_{ij} I_{ij}$ is the dipole interaction energy for the transition $|i\rangle \leftrightarrow |j\rangle$ under the influence of the field in the transmission line with the actual current given by ${\rm{Re}}[I_{ij}(0,t)] = I_{ij}\cos{\omega_{ij}^d t}$. Here we assume that our pointlike atom is situated at $x=0$ and the waves $I_{ij}(x,t) = I_{ij}\exp{(ik_{ij}x-i\omega_{ij}^d t)}$ with the wavevector $k_{ij}$ propagate in the transmission line.  The dipole matrix element can be presented in the form  $\phi_{ij} = \zeta_{ij}MI_p$ with the dimensionality of a magnetic flux, where the dimensionless matrix element satisfies the relation $0\leq \zeta_{ij}\leq 1$.

The atomic dynamics is described by the standard master equation $\dot{\rho} = -(i/\hbar)[H,\rho]+L[\rho]$ for the density matrix $\rho = \sum_{i,j}\rho_{ij}|i\rangle\langle j|$
with the Lindblad term $L = \Gamma_{32}\rho_{33}(-\sigma_{33}+\sigma_{22})+\Gamma_{21}\rho_{22}(-\sigma_{22}+\sigma_{11})+\sum_{i\neq j}\gamma_{ij}\rho_{ij}\sigma_{ij}.$
Here $\gamma_{ij} = \gamma_{ji}$ is the damping rate of the off-diagonal terms (dephasing) and $\Gamma_{ij}$ is the relaxation rate between the levels $|i\rangle$ and $|j\rangle$ ($i>j$). In the ladder-type three-level atom $\Gamma_{31}\ll(\Gamma_{32},\Gamma_{21})$ holds. The low temperature condition suggests suppression of the excitation rates $\Gamma_{12}$ = 0, $\Gamma_{23}$ = 0 and $\Gamma_{13}$ = 0.

The atom in 1D open space generates a scattered wave at frequency $\omega_{21}^d$ \cite{RF}
\begin{equation}\label{Isc}
    I_{\rm sc}(x,t)=i\frac{\hbar\Gamma_{21}}{\phi_{21}}\langle\sigma_{12}\rangle e^{ik_{21}|x|-i\omega_{21}^d t},
\end{equation}
where $\langle\sigma_{ij}\rangle= {\tt tr}[\sigma_{ij}\rho] = \rho_{ji}$ can be straightforwardly found in the stationary conditions ($\dot{\rho}=0$). The transmission coefficient, found as a ratio of the resulting current $I_{21}(x,t) + I_{\rm sc}(x,t)$ at $x>0$ to the incident one $I_{21}(x,t)$, takes the form $t = 1+i(\Gamma_{21}/\Omega_{21})\rho_{21}$, and the amplification condition is $|t| > 1$.

The solution of the master equation \cite{supplementary} is simplified for the most interesting case of nearly resonant drives and fast $|3\rangle\rightarrow|2\rangle$ relaxation ($\Gamma_{32}\gg\Gamma_{21}$), when the state $|3\rangle$ remains nearly unpopulated ($\rho_{33}<\Gamma_{21}/\Gamma_{32} \ll 1$). Neglecting the terms $O(\Gamma_{21}/\Gamma_{32})$, we obtain
\begin{equation}\label{t}
    t \approx 1 + \frac{\Gamma_{21}(\rho_{22}-\rho_{11})}{2\lambda_{21}+\Omega_{31}^2/(2\lambda_{23})},
\end{equation}
where $\lambda_{21}=\gamma_{21}-i\delta\omega_{21}$, $\lambda_{23}=\gamma_{32}+i\delta\omega_{31}-i\delta\omega_{21}$.
In the resonance, this equation gives the standard population inversion condition for the amplification, $\rho_{22}>\rho_{11}$, with the required pumping amplitude
\begin{equation}\label{thramp}
  \Omega_{31} > \sqrt{\Gamma_{21}\Gamma_{32}}
\end{equation}
for weak pure dephasing between the ground and the second excited states ($\gamma_{31} \approx \Gamma_{32}/2$) \cite{supplementary}.


To characterize the energy structure of our atom we perform transmission spectroscopy by sweeping frequency of a weak probe microwave versus the flux bias $\delta\Phi$. The transmitted wave is suppressed due to efficient resonant scattering \cite{RF}, and reveals the transition frequency $\omega_{21}$ as a dark narrow line in $|t|$ on the lower panel of the intensity plot in Fig.~\ref{fig:Spectroscopy}(b). The higher frequency transition between states $|3\rangle$ and $|1\rangle$ cannot be probed in the direct transmission, since the high frequency cutoff of our cryogenic amplifier (13~GHz) is lower than $\omega_{31}$. To detect $\omega_{31}$ we use two-wave spectroscopy in the following way. The probe microwave is adjusted to the minimal transmission $|t_{\rm{min}}|$, taking place at $\omega_{21}^d = \omega_{21}$. Simultaneously we sweep the second high frequency $\omega_{31}^d$. When the transition at $\omega_{31}$ occurs, $|t|$ is increased, as the population of the level $|1\rangle$ is reduced and, therefore, we observe a bright line in the intensity plot $|t/t_{\rm{min}}|$ versus $\delta\Phi$ (the upper panel of Fig.~\ref{fig:Spectroscopy}(b)).

At $\delta\Phi = 0$ the system forms the so-called a ladder type three-level atomic system, in which the selection rule prohibits transitions between levels $|1\rangle$ and $|3\rangle$ due to symmetry of eigenstate wavefunctions. This is seen as the vanishing spectroscopic line in the upper panel of Fig.~\ref{fig:Spectroscopy}(b) at $\omega/2\pi = 34.61$~GHz. To achieve population inversion, we must be able to pump the atom from the ground ($|1\rangle$) to the second excited state ($|3\rangle$), therefore, we choose our working point at $\delta\Phi/\Phi_0 = 3.5\times10^{-3}$, slightly away from $\delta\Phi=0$. Driving the system at $\omega_{31}/2\pi$ = 35.11~GHz, we expect the cascade relaxation $|3\rangle\rightarrow|2\rangle\rightarrow|1\rangle$, accompanied by photon emission at frequencies $\omega_{32}=24.15$~GHz and $\omega_{21} = 10.96$~GHz. The lowest transition can be detected, since it is within our cryogenic amplifier frequency band.

\begin{figure}[tbp]
\includegraphics[width=8.6cm]{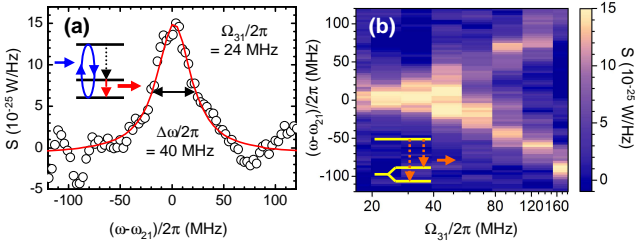}
\caption{Spontaneous emission in the three-level atom.
(a) Emission spectrum ($S = 2\pi S(\omega)$) measured in the vicinity of $\omega_{21}/2\pi$ = 10.96~GHz ($\Phi = 3.5\times10^{-3} \Phi_0$, see Fig. \ref{fig:Spectroscopy}(b)). (b) Emission spectrum as a function of pumping amplitude $\Omega_{31}$. At strong pumping, the emission peak is split by $\Omega_{31}$, due to Rabi splitting of the level $|1\rangle$, as schematically depicted in the inset. The spontaneous emission spectrum gives the noise in the system.}\label{fig:Emission}
\end{figure}

Figure \ref{fig:Emission}(a) shows the measured spontaneous emission spectrum under the resonant pumping at $\omega_{31}$ with an amplitude $\Omega_{31}/2\pi =$ 24~MHz, which confirms that the mechanism of the level $|2\rangle$ population is implemented. At a weak pumping amplitude ($\Omega_{31}/2 \ll (\gamma_{32},\gamma_{21})$), the state $|2\rangle$ population $\rho_{22}$ together with the spontaneous emission rate $\Gamma_{21}$ define the spectral density of the emission in one of the two possible directions via
\begin{equation}\label{Sw}
S(\omega)\approx\frac{\rho_{22}\hbar\omega_{21}\Gamma_{21}}{2\pi}
\frac{\gamma_{21}}{\gamma_{21}^2+(\omega-\omega_{21})^2}.
\end{equation}
Although in our case $\Omega_{31}/2$ is not negligible comparing $\gamma_{21}$, the linewidth $\Delta\omega \approx 2\gamma_{21}$ is still mainly determined by $\gamma_{21}/2\pi \approx$ 18~MHz \cite{supplementary}. The intensity plot in Fig.~\ref{fig:Emission}(b) shows the spontaneous emission from the atom at different pumping amplitudes $\Omega_{31}$. With increasing $\Omega_{31}$, the spontaneous emission peak broadens and then splits due to splitting of the ground state $|1\rangle$ by the Rabi frequency $\Omega_{31}$. The emission in the more general case is calculated analytically in \cite{supplementary} and well describes the split emission spectral line. Particularly, in the extreme case of $\Omega_{31} \gg (\gamma_{32},\gamma_{21})$ each of the two split peaks is expressed as
\begin{equation}\label{Sstrong}
S^\pm(\omega)\approx\frac{\rho_{22}\hbar\omega_{21}\Gamma_{21}}{2\pi}\frac{\gamma'/2}{\gamma'^2 + (\omega-\omega_{21} \pm \Omega_{31}/2)^2},
\end{equation}
where $\gamma' = (\gamma_{32}+\gamma_{21})/2$.

\begin{figure}[tbp]
\includegraphics[width=8.6cm]{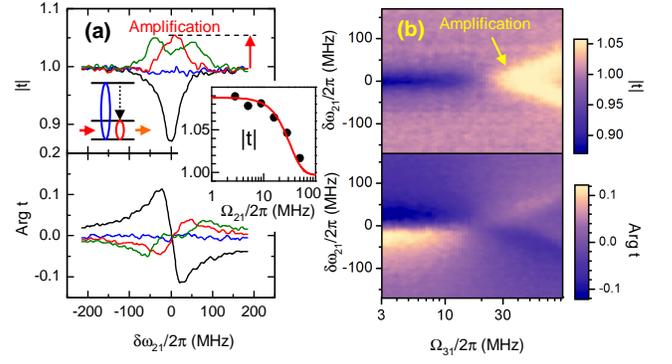}
\caption{Amplification on the single three-level atom.
(a) Amplitude and phase of the transmission coefficient $t$ versus detuning from the resonant frequency $\omega_{21}/2\pi$ = 10.96~GHz at a different pumping amplitudes $\Omega_{31}$. The inset shows the amplification coefficient $|t|$ as a function of the probe amplitude $\Omega_{21}$ at $\delta\omega_{21} = 0$.
(b) The intensity plot, summarizing the measurements shown in (a), demonstrates the transmission amplitude (upper panel) and phase (lower panel) as a function of the pumping amplitude $\Omega_{31}$.}\label{fig:Amplification}
\end{figure}

Equations~(\ref{Sw}) and (\ref{Sstrong}) describe spectrum of the quantum noise in the system, determined by the spontaneous emission to the transmission line. The interaction with 1D open space (transmission line) characterized by $\Gamma_{21}$ is enhanced in the field of an external resonant probe wave at the transition frequency $\omega_{21}$, stimulating emission coherent with the external probe wave. We measure the amplitude and phase of transmission $t$ as a function of detuning $\delta\omega_{21}$ under a relatively weak probe wave amplitude $\Omega_{21}$, at different pumping amplitudes $\Omega_{31}$ (Fig.~\ref{fig:Amplification}(a)). The black curve, obtained at relatively weak pumping with $\Omega_{31}/2\pi$ = 3~MHz, shows a Lorentzian dip with the linewidth of 40~MHz, determined mainly by the dephasing $2\gamma_{21}$. The transmission is strongly modified as $\Omega_{31}$ is increased. At $\Omega_{31}/2\pi$ = 23~MHz, the dip is completely suppressed (blue curve). At stronger pumping, the amplification is observed: at $\Omega_{31}/2\pi$ = 40~MHz (red curve) the transmission exceeds one by 5\%, exhibiting a clear amplification peak, and at $\Omega_{31}/2\pi$ = 95~MHz (green curve) the peak is split. Note that in the amplification condition, the phase on the lower panel of Fig.~\ref{fig:Amplification}(a) is inverted. The intensity plots of Fig.~\ref{fig:Amplification}(b) summarize the  behavior of $t$ versus pumping amplitude $\Omega_{31}$.

The inset of Fig.~\ref{fig:Amplification}(a) shows gain $|t|$ at the fixed pumping amplitude $\Omega_{31}/2\pi$ = 40~MHz as a function of the probe amplitude $\Omega_{21}$. In the linear amplification regime ($\Omega_{21}/2\pi <$ 20~MHz), corresponding to the nearly constant $|t|$, the gain reaches maximum of about 1.09. Interestingly, the best agreement between the calculated (red) curve and the experimental data (black dots) is obtained when pure dephasing is neglected ($\gamma_{21} = \Gamma_{21}/2$). Indeed, $t$ is insensitive to linear fluctuations of $\delta\omega_{21}$ in the ``magic point" of $\Omega_{31} = 2\gamma_{32}$.

Remarkably, the gain for the single atom amplifier is fundamentally limited by $\sqrt 2$ (the corresponding power gain is $2$), as each incident photon can stimulate not more than one emitted photon. Moreover, the photon multiplication factor is always less than 2, due to competing incoherent processes, for instance, relaxation caused by unavoidable quantum noise. The factor is reduced even more by less-than-100\% population of the level $|2\rangle$.
To calculate the maximal gain we consider the system in double resonance, without pure dephasing and with $\Gamma_{32} \gg \Gamma_{21}$. In such a case for driving amplitude of the order of threshold defined by Eq.~(\ref{thramp}), the populations can be simplified to $\rho_{11} = 1/(1 + \nu)$ and $\rho_{22} = \nu/(1 + \nu)$, where $\nu = \Omega_{31}^2/(\Gamma_{32}\Gamma_{21})$ is a square of normalized driving amplitude \cite{supplementary}. With this variable Eq. (\ref{t}) becomes $t = 1 + (\nu - 1)/(\nu + 1)^2$, which takes maximum $t = 1 + 1/8$ at $\nu = 3$, that is $\Omega_{31}^2 = 3\Gamma_{21}\Gamma_{32}$ and the corresponding populations are $\rho_{22} = 3/4$ and $\rho_{11} = 1/4$.

We derive the parameters of our system from the experimental data. From the dip in the transmission in the absence of pumping (see Eq. (\ref{t}) with $\Omega_{31} = 0$ and $\rho_{22}-\rho_{11} = -1$), $\Gamma_{21}/2\pi$ is found to be 11~MHz. In the case of negligible pure dephasing in $\gamma_{31}$ ($= \Gamma_{32}/2)$ and $\gamma_{32}$ ($= \Gamma_{32}/2 + \Gamma_{21}/2$) \cite{puredeph3}, the experimental data are in a very good agreement with theory, when we take $\Gamma_{32}/2\pi$ = 35~MHz \cite{G32}. Amplification occurs, when the pumping amplitude exceeds $\Omega_{31}/2\pi \approx$ 20~MHz, found from Eq.~(\ref{thramp})  consistently with our experiment.

\begin{figure}[tbp]
\includegraphics[width=8.6cm]{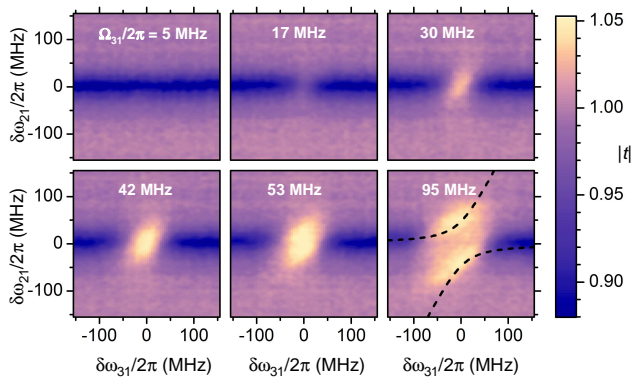}
\caption{Amplification versus detuning of pumping and probe frequencies $\delta\omega_{32}$ and $\delta\omega_{21}$. The pumping amplitudes $\Omega_{31}$ are indicated on the panels. The black dashed line on the panel with the strongest pumping ($\Omega_{31}/2\pi$ = 95~MHz) shows the expected anticrossing spectroscopy line.}\label{fig:Detuning}
\end{figure}

The six panels in Fig.~\ref{fig:Detuning}, each measured at different $\Omega_{31}$, show $|t|$ versus detuning from the resonances: $\delta\omega_{31}$ ($x$-axis) and $\delta\omega_{21}$ ($y$-axis). Amplification regions revealed as bright spots near the double resonance points, split at strong pumping ($\Omega_{31}/2\pi$ = 95~MHz). The splitting is reminiscent of the typical anticrossing (shown by a black dashed line); however, it reveals in the over-unity transmission $|t|>1$ that is amplification.

The amplification by a single artificial atom provides an example of an elementary quantum amplifier. However, we suggest that it can be used as a building block for practical quantum amplifiers with the noise characteristics limited by quantum noise due to spontaneous emission to the open space. The demonstrated on-chip amplifier is tunable, and its bandwidth can be selected by changing coupling to the transmission line.

In a conclusion, we have demonstrated a fully controllable and tunable on-chip quantum amplifier with a single artificial atom in open 1D space. The work may open a direction of on-chip quantum electronics.

This work was supported by CREST-JST and MEXT kakenhi ``Quantum Cybernetics".


\newpage

\newcommand{\tr}{\color{red}}
\newcommand{\tb}{\color{black}}


\def\theequation{S\arabic{equation}}
\setcounter{equation}{0}

\def\thefigure{S\arabic{figure}}
\setcounter{figure}{0}
\begin{widetext}
\part*{Supplementary materials:\\ Ultimate on-chip quantum amplifier}
\end{widetext}

\renewcommand{\thesection}{\Roman{section}}
\renewcommand{\thesubsection}{\Alph{subsection}}

\section{Three-level atom dynamics}

\subsection{Population inversion and amplification}

In the rotating wave approximation, the three-level system  is described by the Hamiltonian
\begin{equation}\label{SEq1}
    H_a = -\hbar(\delta\omega_{21} \sigma_{22} + \delta\omega_{31} \sigma_{33}),
\end{equation}
where $\sigma_{ij} = |i\rangle\langle j|$ is the atomic projection/transition operator, $\delta\omega_{ij}=\omega_{ij}^d-\omega_{ij}$.
 The external driving microwave fields at frequencies $\omega_{31}^d$ and $\omega_{21}^d$ couple atomic states according to the interaction Hamiltonian
\begin{equation}\label{SEq2}
    H_{int} = -\hbar\Bigg[\frac{\Omega_{31}}{2}(\sigma_{31}+\sigma_{13}) + \frac{\Omega_{21}}{2}(\sigma_{21}+\sigma_{12})\Bigg],
\end{equation}
 where $\hbar\Omega_{ij} = \phi_{ij} I_{ij}$ is the dipole interaction energy for the transition $|i\rangle \leftrightarrow |j\rangle$ under the influence of the field in the transmission line with the actual current given by ${\rm{Re}}[I_{ij}(0,t)] = I_{ij}\cos{\omega_{ij}^d t}$. Here we assume that our pointlike atom is situated at $x=0$ and the waves from microwave sources $I_{ij}(x,t) = I_{ij}\exp{(ik_{ij}x-i\omega_{ij}^d t)}$ (with the wavevector $k_{ij}$) propagate in the transmission line.  The dipole matrix element can be presented in the form  $\phi_{ij} = \zeta_{ij}MI_p$ with the dimensionality of a magnetic flux, where the dimensionless matrix element $\zeta_{ij}$ satisfies the condition of $0\leq \zeta_{ij}\leq 1$,  $M$ is the line-atom mutual inductance, and $I_p$ is the amplitude of the persistent current in the loop.

The atomic dynamics is described by the standard master equation for the density matrix $\rho = \sum_{i,j}\rho_{ij}|i\rangle\langle j|$,
\begin{equation}\label{Master}
\dot{\rho} = -(i/\hbar)[H,\rho]+L[\rho]
\end{equation}
with the Lindblad term
\begin{equation}\label{master}
    L = \Gamma_{32}\rho_{33}(-\sigma_{33}+\sigma_{22})+\Gamma_{21}\rho_{22}(-\sigma_{22}+\sigma_{11})+\sum_{i\neq j}\gamma_{ij}\rho_{ij}\sigma_{ij}.
\end{equation}
Here $\gamma_{ij} = \gamma_{ji}$ is the damping rate of the off-diagonal terms (dephasing) and $\Gamma_{ij}$ is the relaxation rate between the levels $|i\rangle$ and $|j\rangle$ ($i>j$). In the ladder-type three-level atom, the $|3\rangle\rightarrow|1\rangle$ transitions can be neglected since $\Gamma_{31}\ll(\Gamma_{32},\Gamma_{21})$. Our experiment is performed at the temperature $T$, low compared to the atomic energy splitting ($\hbar\omega_{ij}\gg k_BT$), which guarantees the absence of thermal excitations ($\Gamma_{12}$ = 0, $\Gamma_{23}$ = 0 and $\Gamma_{13}$ = 0).

The atom interacting only with continuum modes of 1D open space generates a scattered wave \cite{RF}
\begin{equation}\label{SIsc}
    I_{\rm sc}(x,t)=i\frac{\hbar\Gamma_{21}}{\phi_{21}}\langle\sigma_{12}\rangle e^{ik_{21}|x|-i\omega_{21} t},
\end{equation}
where $\langle\sigma_{ij}\rangle= {\tt tr}[\sigma_{ij}\rho] = \rho_{ji}$ can be straightforwardly found in the stationary conditions ($\dot{\rho}=0$), when the master equation reduces to a set of linear algebraic equations. The transmission coefficient found as a  ratio of the resulting current $I_{21}(x,t) + I_{\rm sc}(x,t)$ (at $x>0$) to the incident one $I_{21}(x,t)$ takes the form
\begin{equation}\label{St}
    t = 1+i\frac{\Gamma_{21}}{\Omega_{21}}\rho_{21}.
\end{equation}
From Eq.~(\ref{St}), the amplification condition is
\begin{equation}
    |t| > 1.
\end{equation}

In the case of weak drive (probe) at around $\omega_{21}$, when it does not change the population of levels, solutions for the density matrix elements are
\begin{equation}\label{rho11}
    \rho_{11} = \frac{A}{A+B+1},
\end{equation}
\begin{equation}\label{rho22}
    \rho_{22} = \frac{B}{A+B+1},
\end{equation}
\begin{equation}\label{rho33}
    \rho_{33} = \frac{1}{A+B+1},
\end{equation}
\begin{equation}\label{rho21}
    \rho_{21} = \frac{-i\frac{\Omega_{21}}{2\lambda_{21}}(\frac{\Omega_{31}^2}{4\lambda_{13}\lambda_{23}}(\rho_{11}-\rho_{33})+\rho_{22}-\rho_{11})}{1+\frac{\Omega_{31}^2}{4\lambda_{21}\lambda_{23}}},
\end{equation}
where $\lambda_{ij} = \lambda_{ji}^*$, $\lambda_{21}=\gamma_{21}-i\delta\omega_{21}$, $\lambda_{23}=\gamma_{32}+i\delta\omega_{31}-i\delta\omega_{21}$, $\lambda_{13}=\gamma_{31}+i\delta\omega_{31}$,  $A=2\Gamma_{32}|\lambda_{13}|^2/(\gamma_{31}\Omega_{31}^2)+1$, $B=\Gamma_{32}/\Gamma_{21}$.

Equation~(\ref{rho21}) simplifies for the most interesting case of nearly resonant drives and fast $|3\rangle\rightarrow|2\rangle$ relaxation ($\Gamma_{32}\gg\Gamma_{21}$), where the state $|3\rangle$ remains nearly unpopulated ($\rho_{33}<\Gamma_{21}/\Gamma_{32} \ll 1$). Neglecting the terms   $O(\Gamma_{21}/\Gamma_{32})$, we arrive at
\begin{equation}\label{rho21s}
    \rho_{21} \approx -i\frac{\Omega_{21}(\rho_{22}-\rho_{11})}{2\lambda_{21}+\frac{\Omega_{31}^2}{2\lambda_{23}}}.
\end{equation}
from which, according to Eq.~(\ref{t}),
\begin{equation}\label{tt}
   t \approx 1 + \frac{\Gamma_{21}(\rho_{22}-\rho_{11})}{2\lambda_{21}+\frac{\Omega_{31}^2}{2\lambda_{23}}}
\end{equation}
This equation gives the standard population inversion condition for the amplification
\begin{equation}
    \rho_{22}>\rho_{11},
\end{equation}
with the required pumping amplitude
\begin{equation}
    \Omega_{31}^2 > 2\Gamma_{21}\gamma_{31}.
\end{equation}
For weak pure dephasing between the ground and the second excited states, the total dephasing is simplified to $\gamma_{31} \approx \Gamma_{32}/2$ and the amplification condition takes the form
\begin{equation}\label{OmegaAmp}
    \Omega_{31}^2 > \Gamma_{21}\Gamma_{32}.
\end{equation}

One interesting consequence of Eq.~(\ref{tt}) is that $t$ becomes insensitive to low frequency fluctuations between $|1\rangle$ and $|2\rangle$ at some optimal driving amplitude $\Omega_{31}$. This ``magic" point, where the pure dephasing is suppressed ($\gamma_{21} = \Gamma_{21}/2$), takes  place at $\Omega_{31} = 2\gamma_{32}$ because the denominator of $t$ in Eq.~(\ref{t}) contains a linear term $i \delta\omega_{21} (\Omega_{31}^2/4\gamma_{32} - 1)$, while its numerator does not depend on $\delta\omega_{21}$.

The interesting question: What is the maximal possible gain of the single-atom amplifier? We consider the ideal case of absence of pure dephasing [$\gamma_{21} = \Gamma_{21}/2$, $\gamma_{32} = \Gamma_{32}/2$, $\gamma_{31} = \Gamma_{32}/2$ (because $\Gamma_{31}\ll\Gamma_{21}\ll\Gamma_{32}$)] $|t|$ from Eq.~(\ref{tt}) in the double resonance $\delta\omega_{21} = 0$ and $\delta\omega_{31} = 0$. When the pumping amplitude $\Omega_{31}$ is of order of the threshold defined by Eq.~(\ref{OmegaAmp}), the atomic populations can be expressed as $\rho_{11} = 1/(1 + \nu)$ and $\rho_{22} = \nu/(1 + \nu)$, where $\nu = \Omega_{31}^2/(\Gamma_{32}\Gamma_{21})$ is the square of the normalized pumping. The transmission coefficient $t = 1 + (\nu - 1)/(1 + \nu)^2$ reaches its maximum at $\nu = 3$, that is
\begin{equation}
\Omega_{31}^2 = 3\Gamma_{21}\Gamma_{32},
\end{equation}
which corresponds to the gain
\begin{equation}
    |t| = 1\frac{1}{8}.
\end{equation}

\subsection{Spectrum of spontaneous emission}

The spontaneous emission spectrum of the on-chip single atom is affected by a strong pumping $\Omega_{31}$. We find the emission power spectral density of the waves, propagated in one direction (forward or backward), for $|2\rangle\rightarrow|1\rangle$ transition \cite{Scully}
\begin{equation}
S(\omega)=\frac{\hbar\omega\Gamma_{21}}{2\pi}{\rm{Re}}\int^\infty_{0}{\langle\sigma_{21}(0)\sigma_{12}(\tau)\rangle e^{i\omega\tau}d\tau}.
\end{equation}
The fact that only half power is detected is accounted as a 1/2 factor in the equation.

In the condition of resonant pumping at $\omega_{31}$, we find a form of Eq.~(\ref{Master})
\begin{equation}
\frac{d}{dt}
\left(
  \begin{array}{c}
    \rho_{21} \\
    \rho_{23} \\
  \end{array}
\right)
=
\left(
  \begin{array}{cc}
    -\gamma_{21} & -i\Omega_{31}/2 \\
    -i\Omega_{31}/2 & -\gamma_{32} \\
  \end{array}
\right)
\left(
  \begin{array}{c}
    \rho_{21} \\
    \rho_{23} \\
  \end{array}
\right).
\end{equation}
Using the quantum regression theorem, we obtain from here the   following equation for two-time correlation functions
\begin{equation}\label{twotimes}
\frac{d\mathbf{s}}{d\tau} = M \mathbf{s},
\end{equation}
where
\begin{equation}
\mathbf{s} =
\left(
  \begin{array}{cc}
    \langle\sigma_{21}(0)\sigma_{12}(\tau)\rangle \\
    \langle\sigma_{21}(0)\sigma_{32}(\tau)\rangle \\
  \end{array}
\right),
\end{equation}
\begin{equation}
M=
\left(
  \begin{array}{cc}
    -\gamma_{21} & -i\Omega_{31}/2 \\
    -i\Omega_{31}/2 & -\gamma_{32} \\
  \end{array}
\right)
\end{equation}
with the initial conditions defined by the steady state
\begin{equation}
\mathbf{s_{s}} =
\left(
  \begin{array}{cc}
    \langle\sigma_{22}\rangle \\
    0 \\
  \end{array}
\right).
\end{equation}
In the absence of driving at $\omega_{21}$ and $\omega_{32}$ the steady states give $\langle\sigma_{12}\rangle_s=0$ and $\langle\sigma_{32}\rangle_s=0$, therefore $\sigma_{12}(t)$ and $\sigma_{32}(t)$ present incoherent dynamics (fluctuations).
Solving Eq.~(\ref{twotimes}) and returning to the laboratory frame, we find for $\tau > 0$
\begin{equation}
\langle \sigma_{21}(0)\sigma_{12}(\tau)\rangle=\langle \sigma_{22}\rangle \frac{\sinh(\Lambda\tau/2+\alpha)}{\sinh{\alpha}}e^{(-\gamma/2-i\omega_{21})\tau},
\end{equation}
where $\Lambda=\sqrt{\Delta\gamma^2-\Omega_{31}^2}$, $\Delta\gamma=(\gamma_{32}-\gamma_{21})$, $\gamma=\gamma_{32}+\gamma_{21}$, $\sinh{\alpha}=\Lambda/\Omega_{31}$. The emission spectrum is
\begin{widetext}
\begin{equation}\label{Sweak}
\nonumber S(\omega)=\frac{\rho_{22}\hbar\omega_{21}\Gamma_{21}}{2\pi}
\Bigg[
\frac{(\gamma-\Lambda)(\Delta\gamma/\Lambda+1)}{(\gamma-\Lambda)^2+4\delta\omega^2}-
\frac{(\gamma+\Lambda)(\Delta\gamma/\Lambda-1)}{(\gamma+\Lambda)^2+4\delta\omega^2}
\Bigg]
\end{equation}
in the case of $\Omega_{31}< |\Delta\gamma|$, where $\delta\omega=\omega-\omega_{21}$, and
\begin{equation}\label{SSstrong}
S(\omega)=\frac{\rho_{22}\hbar\omega_{21}\Gamma_{21}}{2\pi}\Bigg[
\frac{\gamma+2(\delta\omega+\Omega'/2)\Delta\gamma/\Omega'}{\gamma^2+4(\delta\omega+\Omega'/2)^2}+
\frac{\gamma-2(\delta\omega-\Omega'/2)\Delta\gamma/\Omega'}{\gamma^2+4(\delta\omega-\Omega'/2)^2}
\Bigg]
\end{equation}
in the case of $\Omega_{31}>|\Delta\gamma|$. Here we have introduced a real variable  $\Omega'=-i\Lambda=\sqrt{\Omega_{31}^2-\Delta\gamma^2}$.
\end{widetext}

In the limit of weak pumping $\Omega_{31}\ll (\gamma_{32},\gamma_{21})$, the equations are simplified to the well known form for spontaneous emission of a two-level atom
\begin{equation}\label{Sweak2}
S(\omega) \approx \frac{\rho_{22}\hbar\omega_{21}\Gamma_{21}}{2\pi}
\frac{\gamma_{21}}{\gamma_{21}^2+\delta\omega^2}.
\end{equation}
In the strong pumping limit $\Omega_{31}\gg\gamma$, the emission spectrum
\begin{widetext}
\begin{equation}\label{SSstrong2}
S(\omega)\approx\frac{\rho_{22}\hbar\omega_{21}\Gamma_{21}}{2\pi}\Bigg[
\frac{\gamma}{\gamma^2+4(\delta\omega+\Omega_{31}/2)^2}+
\frac{\gamma}{\gamma^2+4(\delta\omega-\Omega_{31}/2)^2}
\Bigg]
\end{equation}
\end{widetext}

is split into two peaks by frequency $\Omega_{31}$ and width of $\gamma$. The splitting can be explained in the rotating frame by relaxation from the level $|2\rangle$ to the split level $|1\rangle$.

\section{Material and Methods}

\subsection{Sample design and fabrication}

The coplanar transmission line with the characteristic impedance $Z\simeq50\,\Omega$ is made by patterning a gold film deposited on a silicon substrate. In the middle of the chip, the central conductor of the waveguide is narrowed and replaced by aluminium. The latter is deposited together with the artificial atom of the flux qubit geometry shown in Fig.~\ref{fig:Setup}(a), using shadow evaporation. The experiment is performed in a dilution refrigerator at a temperature of 40 mK.

\subsection{Transmission and reflection coefficients of the elastic
scattering}

We measure the complex transmission coefficient $t$ by a vector network analyzer. The transmission characteristics of the microwave line are calibrated with the artificial atom effectively removed: It is offset by a dc flux bias such that its transition frequency does not fall in the measurement frequency range. A schematic diagram of the experimental setup for measuring the transmission coefficient is presented in Fig.~\ref{fig:Setup}(b).

\subsection{Measurements of the spontaneous emission spectrum}

\begin{figure}[hbt]
\includegraphics[width=10cm]{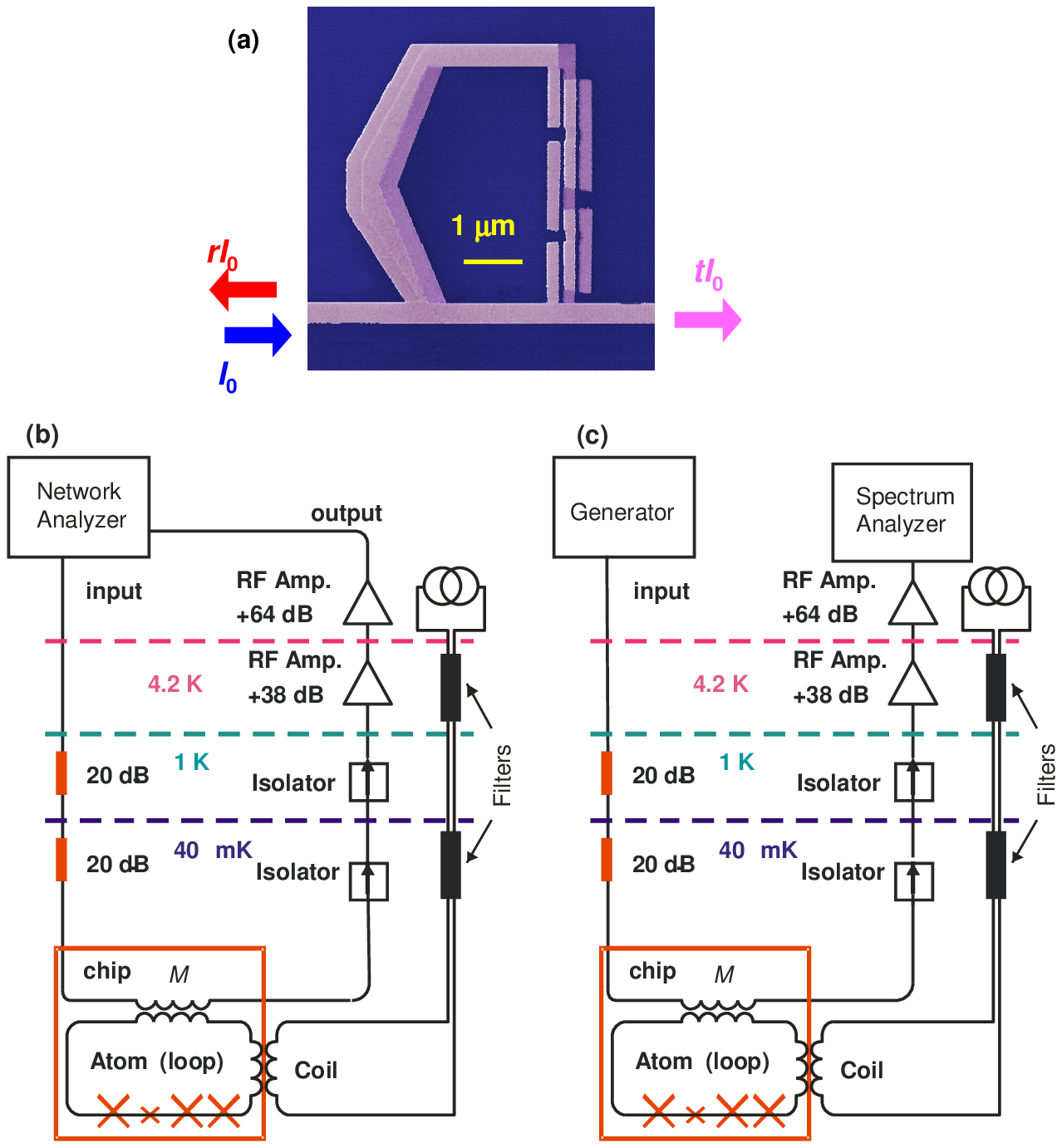}
\caption{(a) False colored SEM image of the artificial atom. The ``atomic" loop of is coupled to the line via mutual inductance of the shared segment with the transmission line. Reflection and transmission amplitudes ($r$ and $t$) can be monitored. (b) Experimental setup diagram for measuring amplification (transmission). To measure the transmission coefficient $t$ the input signal from a network analyzer is applied to the channels ``input", and amplitude and phase of an output signal are detected. The input line contains 20~dB attenuator at 1~K stage and another 20~dB attenuator at the base temperature to suppress 300~K blackbody radiation. The weak signal passing through the sample is then amplified using a cryogenic amplifier at 4.2~K and a room temperature amplifier. For suppression blackbody radiation coming back from the 4~K amplifier, we use two isolators at the base temperature and at the 1~K stage providing 40~dB total attenuation for the backward signal. The excitation energies of the artificial atom are controlled using a coil magnetic field. (c) Experimental setup for measuring the spectrum of spontaneous emission. We pump the atom at a high frequency corresponding transition between the ground and the second excited state and measure the emission spectrum of the first excited to the ground state using a spectrum analyzer. The weak atomic emission spectrum is extracted by subtraction of spectral curves with ``on" ``off" excitation power.}\label{fig:Setup}
\end{figure}

To measure weak spontaneous emission on the background of top of the amplifier noise, we modulate the excitation power and use digital differentiation. The noise spectral density of the preamplifier situated at 4 K is $2\pi S(\omega) = k_B T_n\simeq 1.9 \times 10^{-22}$~W/Hz, which corresponds to the effective noise temperature of the amplifier at the input $T_n = 14$~K. A schematic diagram of the setup for measuring the spontaneous emission spectrum is presented in Fig.~\ref{fig:Setup}(c).

\subsection{Calibration and calculations}

The parameters of our device are found from the experimental curves in the following way. For calibration of the probe power at $\omega_{21}$ we measure dependence of transmission coefficients versus the probe power in the absence of pumping \cite{RF}. For the pumping power calibration, we analyzed splittings of peaks in Fig.~2(b) and Fig.~3(b). The dephasing and relaxation rates are found to be $\gamma_{21}/2\pi = 18$ MHz and $\Gamma_{21}/2\pi = 11$ MHz, correspondingly,  from the dip in transmission in the limit of a weak probing $\Omega_{21}$ (similar to the black curve on Fig. 3(a), but with a weaker probe). The dip width is equal to $2\gamma_{21}$ and the dip amplitude (minimal transmission) is $t_{\rm min} = 1 - \Gamma_{21}/(2\gamma_{21})$. The relaxation rate $\Gamma_{32}$ was adjusted in the simulations to have a better agreement with the experimental curves (particularly, the curves of Fig.~3(a)) with neglected dephasing rates in $\gamma_{31}$ and $\gamma_{32}$: we take $\gamma_{31} = \Gamma_{32}/2$ and $\gamma_{32} = \Gamma_{32}/2+\Gamma_{21}/2$. The pure dephasing rates in $\gamma_{31}$ and $\gamma_{32}$ was neglected as they are expected to be weaker than that in $\gamma_{21}$, due to two times less steep energy bands ($\hbar\omega_{31}$ and $\hbar\omega_{32}$), in assumption that it is a result of the low frequency noise in flux degree of freedom. $\Gamma_{32}/2\pi$ was found to be 35 MHz.

\begin{figure}[tbph]
\includegraphics[width=8cm]{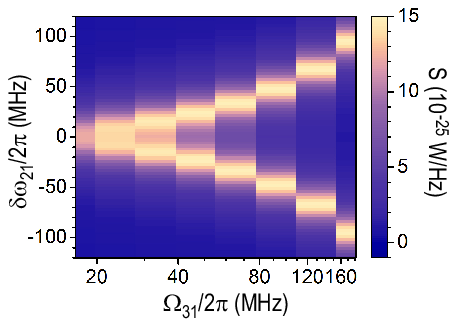}
\caption{Calculated spontaneous emission according to Eqs.~(\ref{Sweak}), (\ref{SSstrong}) for the conditions of Fig. 2(b).}\label{fig:SEmission}
\end{figure}

\begin{figure}[hbt]
\includegraphics[width=8cm]{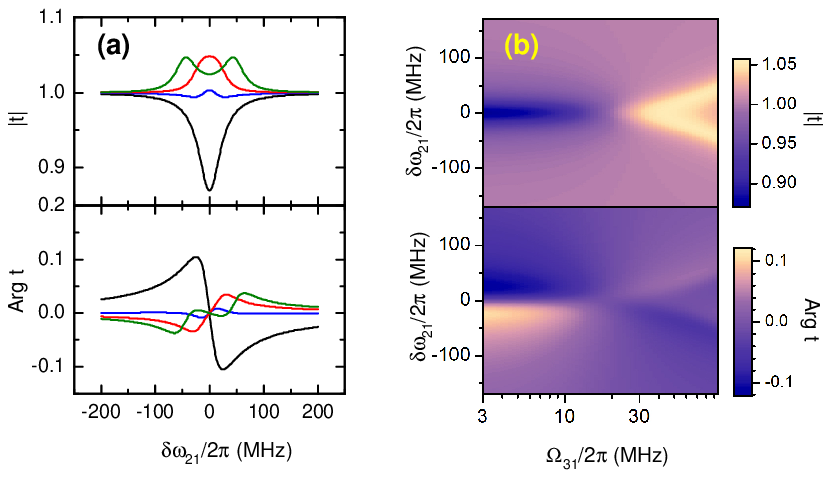}
\caption{Amplification as a function of driving amplitudes. (a) Calculated transmission amplitude and phase for the same pumping amplitudes $\Omega_{31}$ as presented in Fig.~3(a).  (b) Calculated transmission amplitude and phase as a function of the pumping amplitude $\Omega_{31}$ and the probe detuning $\delta \omega_{21}$ for the same conditions as in Fig. 3(b).}\label{fig:Amp}
\end{figure}

\begin{figure}[tbp]
\includegraphics[width=8cm]{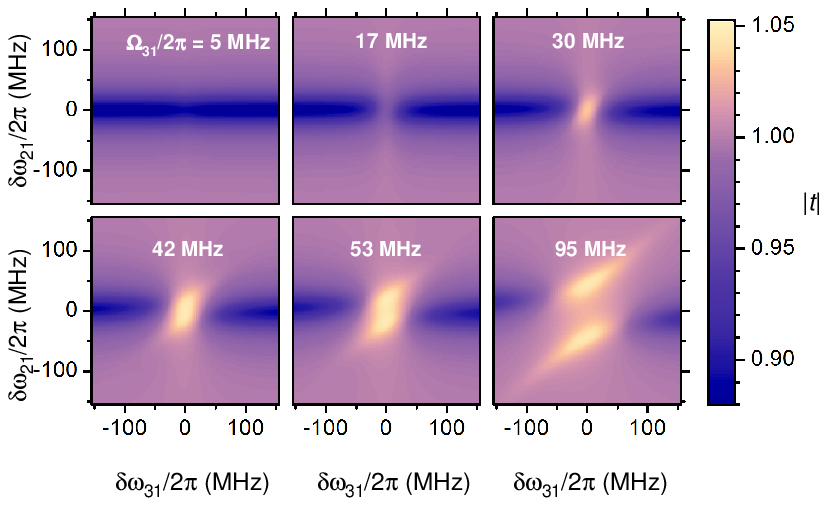}
\caption{Amplification versus detuning of pumping and probing frequencies. Calculated transmission amplitude $|t|$ for different pumping amplitudes $\Omega_{31}$. The conditions are taken the same as for Fig.~4.}\label{fig:2D}
\end{figure}

The spontaneous emission spectrum calculated from Eqs.~(\ref{Sweak}) and (\ref{SSstrong}) shown in Fig.~\ref{fig:SEmission} well reproduces experimentally measured spectrum of Fig.~2(b). Figures \ref{fig:Amp} and \ref{fig:2D} show calculations of the transmission coefficient with the same values of pumping amplitudes as in Figs.~3(b) and 4, correspondingly. In the latter calculations we used full solution of the master equation (not the weak probing case provided by Eqs. (\ref{rho11})-(\ref{rho21}), because the probe amplitude $\Omega_{21}/2\pi =$ 16 MHz is not negligible comparing to $\gamma_{21}$ and $\Gamma_{21}$.


\begin{thebibliography}{27}
\bibitem[{pur()}]{puredeph3}
\bibinfo{note}{We can neglect pure dephasing in $\gamma_{32}$ and
  $\gamma_{31}$, because, first, it is expected to be much weaker than
  $\gamma_{21}$, due to less steep energy bands, and, second, relaxation
  $\Gamma_{32}$ is strong ($\Gamma_{32} \gg \gamma_{21}$). Note also that
  $\Gamma_{31} \ll (\Gamma_{32},\Gamma_{21})$.}
\bibitem[{G32()}]{G32}
\bibinfo{note}{The experimentally derived $\Gamma_{32}$ is about twice higher
  than the expected one for 50 $\Omega$ impedance. We suppose that this is a
  signature of frequency dependent impedance above 13 GHz, which we could not
  control in our experiments.}
\end{thebibliography}

\end{document}